\documentclass[a4paper]{llncs}
\usepackage{longtable}
\usepackage{proof}
\usepackage[table]{xcolor}
\usepackage{amsmath}
\usepackage{mathtools}
\usepackage{setspace}
\usepackage{cite}
\usepackage{tikz-qtree,tikz-qtree-compat}
\usepackage{xcolor}
\usepackage{float}
\usepackage{enumitem}
\usepackage{calligra}
\usepackage{graphicx}
\usepackage{algorithm,algpseudocode}

\makeatletter
\newcommand{\algmargin}{\the\ALG@thistlm}
\makeatother
\newlength{\whilewidth}
\algdef{SE}[parWHILE]{parWhile}{EndparWhile}[1]
  {\parbox[t]{\dimexpr\linewidth-\algmargin}{%
     \hangindent\whilewidth\strut\algorithmicwhile\ #1\ \algorithmicdo\strut}}{\algorithmicend\ \algorithmicwhile}%
\algnewcommand{\parState}[1]{\State%
  \parbox[t]{\dimexpr\linewidth-\algmargin}{\strut #1\strut}}
  
\DeclareFontShape{T1}{calligra}{m}{n}{<->s*[2.2]callig15}{}
\DeclareMathAlphabet{\matcalligra}{T1}{calligra}{m}{n}

\DeclareMathAlphabet\mathbfcal{OMS}{cmsy}{b}{n}
\let\mathcal\undefined \DeclareMathAlphabet{\mathcal}{OMS}{cmsy}{m}{n}

\newcommand{\flqsr}{\ensuremath{\mathsf{4LQS^R}}}

\newcommand{\dlssx}{\mathcal{DL}\langle \mathsf{4LQS^{R,\!\times}}\rangle(\D)}
\newcommand{\shdlssx}{\mathcal{DL}_{\D}^{4,\!\times}}
\newcommand{\shdlss}{\mathcal{DL}_{\D}^{4}}
\newcommand{\D}{\mathbf{D}}
\newcommand{\sroiqd}{\mathcal{SROIQ}(\D)}

\newcommand{\defAs}{\coloneqq}
\newcommand{\I}{\mathbf{I}}

\newcommand{\Ra}{\mathbf{R_A}}
\newcommand{\Rd}{\mathbf{R_D}}
\newcommand{\sym}{\mathsf{Sym}}
\newcommand{\asym}{\mathsf{Asym}}
\newcommand{\refl}{\mathsf{Ref}}
\newcommand{\irref}{\mathsf{Irref}}
\newcommand{\tra}{\mathsf{Tra}}
\newcommand{\fun}{\mathsf{Fun}}

\newcommand{\vipcomment}[1]{}
\newcommand{\pow}{\mathcal{P}}

\newcommand{\T}{\mathcal{T}}

\newcommand{\seq}{\mathcal{S}^{\overline{\beta}}_i}

\newcommand{\seqnj}{\mathcal{S}^{\overline{\beta}}_j}
\newcommand{\ke}{KE-tableau}

\newcommand{\KB}{\mathcal{KB}}

\title{A \textsf{C++} reasoner for the description logic $\shdlssx$ \\(Extended Version)\thanks{Work  partially supported by \emph{Gruppo Nazionale per il Calcolo Scientifico (GNCS-INdAM)}.}} 

\author{Domenico Cantone \and Marianna Nicolosi-Asmundo \and \\Daniele Francesco Santamaria}
\institute{
University of Catania, Dept. of Mathematics and Computer Science\\
~email:~\texttt{\{cantone,nicolosi,santamaria\}@dmi.unict.it}
}

\begin{document}
\maketitle

\begin{abstract}
	We present an ongoing implementation of a \ke\space based reasoner for a decidable fragment of stratified elementary set theory expressing the description logic $\dlssx$ (shortly $\shdlssx$). The reasoner checks the consistency of 
	$\shdlssx$-knowledge bases (KBs) represented in set-theoretic terms. It is implemented in \textsf{C++} and supports $\shdlssx$-KBs serialized in the OWL/XML format. 
	
	To the best of our knowledge, this is the first 
	attempt to implement a reasoner for the consistency checking of a description logic represented via a fragment of set theory	that can also classify standard OWL ontologies. 

\end{abstract}

\section{Introduction}

Computable set theory is a research field rich of decidability results, however only recently some of its fragments have been applied in the context of knowledge representation and reasoning for the semantic web. Such efforts are motivated by the characteristics of the considered set-theoretic fragments. These provide very expressive and unique formalisms that combine the modelling capabilities of a rule language with the constructs of description logics. The multi-sorted quantified set-theoretic fragment $\flqsr$ \cite{CanNic2013} is appropriate for these finalities since it turned out to be efficiently implementable. $\flqsr$ involves 
variables of four sorts, pair terms, and a restricted form of quantification over variables of the first three sorts. Its vocabulary contains only the predicate symbols $=$ and $\in$. In spite of that $\flqsr$ allows one to express several constructs of elementary set theory. In particular, is it possible to formalize restricted variants of the set former, which in their turn permit to express other significant set operators such as binary union, intersection, set difference. For example, the powerset of a set $X$, $A = \pow(X)$, is translated into the $\flqsr$-formula $\varphi_1 \equiv  (\forall Z^1)(Z^1 \in A \leftrightarrow (\forall z)(z \in Z^1 \rightarrow z \in X))$, where $z$ is a variable of sort $0$ (individual variable), $Z^1$ and $X$ are variables of sort $1$ (set variables), and $A$ is a variable of sort 2 (collection variable). Within the $\flqsr$ language it is also possible to define binary relations together with several conditions on them which characterize accessibility relations of well known modal logics such as reflexivity and transitivity. For example, a binary relation $R$ is represented by the $\flqsr$-formula $\varphi_2 \equiv (\forall Z^2)(Z^2 \in R \leftrightarrow \neg(\forall z_1)(\forall z_2)\neg (\langle z_1,z_2\rangle = Z^2))$, where $R$ is a variable of sort 3, $Z^2$ is a variable  of sort 2, and $z_1,z_2$ are variables of sort 0. The interested reader may find more examples and details in \cite{CanNic2013}, where decidability of the satisfiability problem for  $\flqsr$ is proved by showing that it enjoys a small model property. In addition, in  \cite{CanNic2013} a family of collections of $\flqsr$-formulae is individuated, each of which having an NP-complete satisfiability problem. It is also shown that the modal logic $\mathsf{K45}$ can be formalized in one of such collections, thus redetermining the NP-completeness of its decision problem \cite{Ladner77thecomputational}.



In \cite{RR2017}, $\flqsr$-quantifier-free atomic formulae of the types $x=y$, $x \in X^1$,  $\langle x,y \rangle \in X^3$ (with $x,y$ variables of sort $0$, $\langle x,y \rangle$ a pair term, $X^1$ a variable of sort $1$, and $X^3$ a variable of sort 3) and $\flqsr$ purely universal formulae of the type $(\forall z_1)...(\forall z_n) \varphi _0$ (with $z_i$ variables of sort $0$, for $i=1,\ldots, n$, and $\varphi _0$ a propositional combination of $\flqsr$-quantifier-free atomic formulae) are used to represent the expressive description logic $\shdlssx$, thus yielding a decision procedure for reasoning tasks for $\shdlssx$ such as the consistency of knowledge bases (KBs) and the \emph{Higher Order Conjunctive Query Answering} problem. The latter problem, in particular, includes the most relevant ABox reasoning tasks.   



The description logic $\shdlssx$ admits full negation, union and intersection of concepts and abstract roles, concept domain and range, existential and minimum cardinality  restriction on the left-hand side of inclusion axioms. It also supports role chains on the left hand side of inclusion axioms and properties on roles such as transitivity, symmetry, reflexivity,
irreflexivity. In some previous work by the authors, the logic  is shown suitable to formalize a rule language such as the Semantic Web Rule Language (SWRL). It has also been shown that, under not very restrictive constraints, its consistency problem is NP-complete. Such a low complexity result is motivated by the fact that existential quantication cannot appear on the right-hand side of inclusion axioms. Nonetheless, $\shdlssx$
turns out to be more expressive than other low complexity
logics such as OWL RL \cite{santaLAP} and therefore it is suitable for representing real world
ontologies. For example, the restricted version of $\shdlssx$ mentioned above allows
one to express several ontologies, such as, for instance, \textsf{OntoCeramic} \cite{cilc15}. Since existential quantification is admitted only on the left hand side of inclusion axioms, $\shdlssx$ is less expressive than logics such as
$\sroiqd\space$ \cite{Horrocks2006} for what concerns the generation of new individuals. On
the other hand, $\shdlssx$ is more liberal than $\sroiqd\space$ in the definition of role
inclusion axioms since the roles involved are not required to be subject to any
ordering relationship, and the notion of simple role is not needed. For example,
the role hierarchy presented in \cite[page~ 2]{Horrocks2006} is not expressible in $\sroiqd\space$
but can be represented in $\shdlssx$. In addition, $\shdlssx$ is a powerful rule language
able to express rules with negated atoms 
that are not supported by the SWRL language.

In this paper we present the first effort to implement 
a \ke\space  based decision procedure for the consistency problem of $\shdlssx$-KBs by resorting to the algorithm introduced in \cite{RR2017}. Implementation is being carried out
in \textsf{C++} 
, as it allows for low level directives and can be easily compiled in several environments. The choice of \ke\space systems \cite{dagostino94} instead of traditional semantic tableaux \cite{smullyan1995first} is motivated by the fact that \ke\space systems introduce an analytic cut rule allowing the construction of trees whose distinct branches define mutually exclusive situations, thus preventing the proliferation of redundant branches, typical of Smullyan's semantic tableaux \cite{smullyan1995first}. Thus, when a consistent KB is given in input, the procedure yields a \ke\space whose open branches induce distinct models of the KB. Otherwise, a closed \ke\space is returned. 



Our reasoner is being developed in Visual Studio 2017  with the compiling tool v.141 for \textsf{C++14} and it is currently in beta-testing phase.  We are also testing it with a virtual machine running Ubuntu with GCC version 4.8.4. The reasoner is available at \texttt{https://github.com/dfsantamaria/DL4xD-Reasoner}.

\section{Preliminaries}

\subsection{The logic $\dlssx$}\label{dlssx}
The description logic $\dlssx$ (which, as already remarked, will be more simply referred to as $\shdlssx$) 
%
admits among other features, Boolean operations on concrete roles, the product of concepts, and also a generic notion of data type, a simple form of concrete domain relevant in real-world applications. In particular, it treats derived data types by admitting data type terms constructed from data ranges by means of a finite number of applications of the Boolean operators. Basic and derived data types can be used inside inclusion axioms involving concrete roles.

Data types are introduced through the notion of data type map, defined  according to \cite{Motik2008} as follows. Let $\D = (N_{D}, N_{C},N_{F},\cdot^{\D})$ be a \emph{data type map}, where  $N_{D}$ is a finite set of data types, $N_{C}$ is a function assigning a set of constants $N_{C}(d)$ to each data type $d \in N_{D}$, $N_{F}$ is a function assigning a set of facets $N_{F}(d)$ to each $d \in N_{D}$, and $\cdot^{\D}$ is a function assigning a data type interpretation $d^{\D}$ to each data type $d \in N_{D}$, a facet interpretation $f^{\D} \subseteq d^{\D}$ to each facet $f \in N_{F}(d)$, and a data value $e_{d}^{\D} \in d^{\D}$ to every constant $e_{d} \in N_{C}(d)$.  We shall assume that the interpretations of the data types in $N_{D}$ are nonempty pairwise disjoint sets.

%
%
%
%

Let $\Ra$, $\Rd$, $\mathbf{C}$, $\mathbf{I}$ be denumerable pairwise disjoint sets of abstract role names, concrete role names, concept names, and individual names, respectively. We assume that the set of abstract role names $\Ra$ contains a name $U$ denoting the universal role. 
%

\vipcomment{An abstract role hierarchy $\mathsf{R}_{a}^{H}$ is a finite collection of RIAs.  A strict partial order $\prec$ on  $\Ra \cup \{ R^- \mid R \in \Ra \}$ is called \emph{a regular order} if $\prec$ satisfies, additionally, $S \prec R$ iff $S^- \prec R$, for all roles R and S.\footnote{We recall that a strict partial order $\prec$  on a set $A$ is an irreflexive and transitive relation on $A$.}}

\noindent
(a) $\shdlssx$-data type, (b) $\shdlssx$-concept, (c) $\shdlssx$-abstract role, and (d) $\shdlssx$-concrete role terms are constructed according to the following syntax rules:
\begin{itemize}
	\item[(a)] $t_1, t_2 \longrightarrow dr ~|~\neg t_1 ~|~t_1 \sqcap t_2 ~|~t_1 \sqcup t_2 ~|~\{e_{d}\}\, ,$
	
	\item[(b)] $C_1, C_ 2 \longrightarrow A ~|~\top ~|~\bot ~|~\neg C_1 ~|~C_1 \sqcup C_2 ~|~C_1 \sqcap C_2 ~|~\{a\} ~|~\exists R.\mathit{Self}| \exists R.\{a\}| \exists P.\{e_{d}\}\, ,$
	
	\item[(c)] $R_1, R_2 \longrightarrow S ~|~U ~|~R_1^{-} ~|~ \neg R_1 ~|~R_1 \sqcup R_2 ~|~R_1 \sqcap R_2 ~|~R_{C_1 |} ~|~R_{|C_1} ~|~R_{C_1 ~|~C_2} ~|~id(C) ~|~ $
	
	$C_1 \times C_2    \, ,$
	
	\item[(d)] $P_1,P_2 \longrightarrow T ~|~\neg P_1 ~|~ P_1 \sqcup P_2 ~|~ P_1 \sqcap P_2  ~|~P_{C_1 |} ~|~P_{|t_1} ~|~P_{C_1 | t_1}\, ,$
\end{itemize}
where $dr$ is a data range for $\D$, $t_1,t_2$ are data type terms, $e_{d}$ is a constant in $N_{C}(d)$, $a$ is an individual name, $A$ is a concept name, $C_1, C_2$ are $\shdlssx$-concept terms, $S$ is an abstract role name,  $R, R_1,R_2$ are $\shdlssx$-abstract role terms, $T$ is a concrete role name, and $P,P_1,P_2$ are $\shdlssx$-concrete role terms. Notice that data type terms are introduced in order to represent derived data types.

A $\shdlssx$-knowledge base is a triple ${\mathcal K} = (\mathcal{R}, \mathcal{T}, \mathcal{A})$ such that $\mathcal{R}$ is a $\shdlssx$-$RBox$, $\mathcal{T}$ is a $\shdlssx$-$TBox$, and $\mathcal{A}$ a $\shdlssx$-$ABox$. 

A $\shdlssx$-$RBox$ is a collection of statements of the following forms: $R_1 \equiv R_2$, $R_1 \sqsubseteq R_2$, $R_1\ldots R_n \sqsubseteq R_{n+1}$, $\sym(R_1)$, $\asym(R_1)$, $\refl(R_1)$, $\irref(R_1)$, $\mathsf{Dis}(R_1,R_2)$, $\tra(R_1)$, $\fun(R_1)$, $R_1 \equiv C_1 \times C_2$, $P_1 \equiv P_2$, $P_1 \sqsubseteq P_2$, $\mathsf{Dis}(P_1,P_2)$, $\fun(P_1)$, where $R_1,R_2$ are $\shdlssx$-abstract role terms, $C_1, C_2$ are $\shdlssx$-abstract concept terms, and $P_1,P_2$ are $\shdlssx$-concrete role terms. Any expression of the type $w \sqsubseteq R$, where $w$ is a finite string of $\shdlssx$-abstract role terms and $R$ is an $\shdlssx$-abstract role term, is called a \emph{role inclusion axiom (RIA)}. 

A $\shdlssx$-$TBox$ is a set of statements of the types:
\begin{itemize}
	\item[-] $C_1 \equiv C_2$, $C_1 \sqsubseteq C_2$, $C_1 \sqsubseteq \forall R_1.C_2$, $\exists R_1.C_1 \sqsubseteq C_2$, $\geq_n\!\! R_1. C_1 \sqsubseteq C_2$, \\$C_1 \sqsubseteq {\leq_n\!\! R_1. C_2}$,
	\item[-] $t_1 \equiv t_2$, $t_1 \sqsubseteq t_2$, $C_1 \sqsubseteq \forall P_1.t_1$, $\exists P_1.t_1 \sqsubseteq C_1$, $\geq_n\!\! P_1. t_1 \sqsubseteq C_1$, $C_1 \sqsubseteq {\leq_n\!\! P_1. t_1}$,
\end{itemize}
where $C_1,C_2$ are $\shdlssx$-concept terms, $t_1,t_2$ data type terms, $R_1$  a $\shdlssx$-abstract role term, $P_1$ a $\shdlssx$-concrete role term. Any statement of the form $C \sqsubseteq D$, with  $C$, $D$ $\shdlss$-concept terms, is a 
\emph{general concept inclusion axiom}.

A $\shdlssx$-$ABox$ is a set of \emph{individual assertions} of the forms: $a : C_1$, $(a,b) : R_1$, 
$a=b$, $a \neq b$, $e_{d} : t_1$, $(a, e_{d}) : P_1$, 
with $C_1$ a $\shdlssx$-concept term, $d$ a data type, $t_1$ a data type term, $R_1$ a $\shdlssx$-abstract role term, $P_1$ a $\shdlssx$-concrete role term, $a,b$ individual names, and $e_{d}$ a constant in $N_{C}(d)$.

The semantics of $\shdlssx$ is given by means of an interpretation $\I= (\Delta^\I, \Delta_{\D}, \cdot^\I)$, where $\Delta^\I$ and $\Delta_{\D}$ are non-empty disjoint domains such that $d^\D\subseteq \Delta_{\D}$, for every $d \in N_{D}$, and
$\cdot^\I$ is an interpretation function.
The definition of the interpretation of concepts and roles, axioms and assertions is illustrated in  Table \ref{semdlss}.
{\small
	\begin{longtable}{|>{\centering}m{2.5cm}|c|>{\centering\arraybackslash}m{6.7cm}|}
		\hline
		Name & Syntax & Semantics \\
		\hline
		
		concept & $A$ & $ A^\I \subseteq \Delta^\I$\\
		
		ab. (resp., cn.) rl. & $R$ (resp., $P$ )& $R^\I \subseteq \Delta^\I \times \Delta^\I$ \hspace*{0.5cm} (resp., $P^\I \subseteq \Delta^\I \times \Delta_\D$)\\
		
		
		individual& $a$& $a^\I \in \Delta^\I$\\
		
		nominal & $\{a\}$ & $\{a\}^\I = \{a^\I \}$\\
		
		dtype  (resp., ng.) & $d$ (resp., $\neg d$)& $ d^\D \subseteq \Delta_\D$ (resp., $\Delta_\D \setminus d^\D $)\\
		
		
		negative data type term & $ \neg t_1 $ & $  (\neg t_1)^{\D} = \Delta_{\D} \setminus t_1^{\D}$ \\
		
		data type terms intersection & $ t_1 \sqcap t_2 $ & $  (t_1 \sqcap t_2)^{\D} = t_1^{\D} \cap t_2^{\D} $ \\
		
		data type terms union & $ t_1 \sqcup t_2 $ & $  (t_1 \sqcup t_2)^{\D} = t_1^{\D} \cup t_2^{\D} $ \\
		
		constant in $N_{C}(d)$ & $ e_{d} $ & $ e_{d}^\D \in d^\D$ \\
		
		
		
		\hline
		data range  & $\{ e_{d_1}, \ldots , e_{d_n} \}$& $\{ e_{d_1}, \ldots , e_{d_n} \}^\D = \{e_{d_1}^\D \} \cup \ldots \cup \{e_{d_n}^\D \} $ \\
		
		data range   &  $\psi_d$ & $\psi_d^\D$\\
		
		data range    & $\neg dr$ &  $\Delta_\D \setminus dr^\D $\\
		
		\hline
		
		top (resp., bot.) & $\top$ (resp., $\bot$ )& $\Delta^\I$  (resp., $\emptyset$)\\
		
		
		negation & $\neg C$ & $(\neg C)^\I = \Delta^\I \setminus C$ \\
		
		conj. (resp., disj.) & $C \sqcap D$ (resp., $C \sqcup D$)& $ (C \sqcap D)^\I = C^\I \cap D^\I$  (resp., $ (C \sqcup D)^\I = C^\I \cup D^\I$)\\
		
		
		valued exist. quantification & $\exists R.{a}$ & $(\exists R.{a})^\I = \{ x \in \Delta^\I : \langle x,a^\I \rangle \in R^\I  \}$ \\
		
		data typed exist. quantif. & $\exists P.{e_{d}}$ & $(\exists P.e_{d})^\I = \{ x \in \Delta^\I : \langle x, e^\D_{d} \rangle \in P^\I  \}$ \\

		self concept & $\exists R.\mathit{Self}$ & $(\exists R.\mathit{Self})^\I = \{ x \in \Delta^\I : \langle x,x \rangle \in R^\I  \}$ \\
		
		nominals & $\{ a_1, \ldots , a_n \}$& $\{ a_1, \ldots , a_n \}^\I = \{a_1^\I \} \cup \ldots \cup \{a_n^\I \} $ \\
		
		\hline
		
		universal role & U & $(U)^\I = \Delta^\I \times \Delta^\I$\\
		
		inverse role & $R^-$ & $(R^-)^\I = \{\langle y,x \rangle  \mid \langle x,y \rangle \in R^\I\}$\\
		
		concept cart. prod. & $ C_1 \times C_2$   &  $ (C_1 \times C_2)^I = C_1^I \times C_2^I$ \\
		
		abstract role complement & $ \neg R $ & $ (\neg R)^\I=(\Delta^\I \times \Delta^\I) \setminus R^\I $\\
		
		abstract role union & $R_1 \sqcup R_2$ & $ (R_1 \sqcup R_2)^\I = R_1^\I \cup R_2^\I $\\
		
		abstract role intersection & $R_1 \sqcap R_2$ & $ (R_1 \sqcap R_2)^\I = R_1^\I \cap R_2^\I $\\
		
		abstract role domain restr. & $R_{C \mid }$ & $ (R_{C \mid })^\I = \{ \langle x,y \rangle \in R^\I : x \in C^\I  \} $\\

		concrete role complement & $ \neg P $ & $ (\neg P)^\I=(\Delta^\I \times \Delta^\D) \setminus P^\I $\\
		
		concrete role union & $P_1 \sqcup P_2$ & $ (P_1 \sqcup P_2)^\I = P_1^\I \cup P_2^\I $\\
		
		concrete role intersection & $P_1 \sqcap P_2$ & $ (P_1 \sqcap P_2)^\I = P_1^\I \cap P_2^\I $\\
		
		concrete role domain restr. & $P_{C \mid }$ & $ (P_{C \mid })^\I = \{ \langle x,y \rangle \in P^\I : x \in C^\I  \} $\\
		
		concrete role range restr. & $P_{ \mid t}$ &  $ (P_{\mid t})^\I = \{ \langle x,y \rangle \in P^\I : y \in t^\D  \} $\\
		
		concrete role restriction & $P_{ C_1 \mid t}$ &  $ (P_{C_1 \mid t})^\I = \{ \langle x,y \rangle \in P^\I : x \in C_1^\I \wedge y \in t^\D  \} $\\
		
		\hline
		
		concept subsum. & $C_1 \sqsubseteq C_2$ & $\I \models_\D C_1 \sqsubseteq C_2 \; \Longleftrightarrow \; C_1^\I \subseteq C_2^\I$ \\
		
		ab. role subsum. & $ R_1 \sqsubseteq R_2$ & $\I \models_\D R_1 \sqsubseteq R_2 \; \Longleftrightarrow \; R_1^\I \subseteq R_2^\I$\\
		
		role incl. axiom & $R_1 \ldots R_n \sqsubseteq R$ & $\I \models_\D R_1 \ldots R_n \sqsubseteq R  \; \Longleftrightarrow \; R_1^\I\circ \ldots \circ R_n^\I \subseteq R^\I$\\
		cn. role subsum. & $ P_1 \sqsubseteq P_2$ & $\I \models_\D P_1 \sqsubseteq P_2 \; \Longleftrightarrow \; P_1^\I \subseteq P_2^\I$\\
		
		\hline
		
		symmetric role & $\sym(R)$ & $\I \models_\D \sym(R) \; \Longleftrightarrow \; (R^-)^\I \subseteq R^\I$\\
		
		asymmetric role & $\asym(R)$ & $\I \models_\D \asym(R) \; \Longleftrightarrow \; R^\I \cap (R^-)^\I = \emptyset $\\
		
		transitive role & $\tra(R)$ & $\I \models_\D \tra(R) \; \Longleftrightarrow \; R^\I \circ R^\I \subseteq R^\I$\\
		
		disj. ab. role & $\mathsf{Dis}(R_1,R_2)$ & $\I \models_\D \mathsf{Dis}(R_1,R_2) \; \Longleftrightarrow \; R_1^\I \cap R_2^\I = \emptyset$\\
		
		reflexive role & $\refl(R)$& $\I \models_\D \refl(R) \; \Longleftrightarrow \; \{ \langle x,x \rangle \mid x \in \Delta^\I\} \subseteq R^\I$\\
		
		irreflexive role & $\irref(R)$& $\I \models_\D \irref(R) \; \Longleftrightarrow \; R^\I \cap \{ \langle x,x \rangle \mid x \in \Delta^\I\} = \emptyset  $\\
		
		func. ab. role & $\fun(R)$ & $\I \models_\D \fun(R) \; \Longleftrightarrow \; (R^{-})^\I \circ R^\I \subseteq  \{ \langle x,x \rangle \mid x \in \Delta^\I\}$  \\
		
		disj. cn. role & $\mathsf{Dis}(P_1,P_2)$ & $\I \models_\D \mathsf{Dis}(P_1,P_2) \; \Longleftrightarrow \; P_1^\I \cap P_2^\I = \emptyset$\\
		
		func. cn. role & $\fun(P)$ & $\I \models_\D \fun(p) \; \Longleftrightarrow \; \langle x,y \rangle \in P^\I \mbox{ and } \langle x,z \rangle \in P^\I \mbox{ imply } y = z$  \\
		
		\hline
		
		data type terms equivalence & $ t_1 \equiv t_2 $ & $ \I \models_{\D} t_1 \equiv t_2 \Longleftrightarrow t_1^{\D} = t_2^{\D}$\\
		
		data type terms diseq. & $ t_1 \not\equiv t_2 $ & $ \I \models_{\D} t_1 \not\equiv t_2 \Longleftrightarrow t_1^{\D} \neq t_2^{\D}$\\
		
		data type terms subsum. & $ t_1 \sqsubseteq t_2 $ &  $ \I \models_{\D} (t_1 \sqsubseteq t_2) \Longleftrightarrow t_1^{\D} \subseteq t_2^{\D} $ \\
		
		\hline
		
		concept assertion & $a : C_1$ & $\I \models_\D a : C_1 \; \Longleftrightarrow \; (a^\I \in C_1^\I) $ \\
		
		agreement & $a=b$ & $\I \models_\D a=b \; \Longleftrightarrow \; a^\I=b^\I$\\
		
		disagreement & $a \neq b$ & $\I \models_\D a \neq b  \; \Longleftrightarrow \; \neg (a^\I = b^\I)$\\
		
		
		ab. role asser. & $ (a,b) : R $ & $\I \models_\D (a,b) : R \; \Longleftrightarrow \;  \langle a^\I , b^\I \rangle \in R^\I$ \\
		
		cn. role asser. & $ (a,e_d) : P $ & $\I \models_\D (a,e_d) : P \; \Longleftrightarrow \;   \langle a^\I , e_d^\D \rangle \in P^\I$ \\

		\hline \caption{Semantics of $\shdlssx$.}\\
		\caption*{\emph{Legenda.} ab: abstract, cn.: concrete, rl.: role, ind.: individual, d. cs.: data type constant, dtype: data type, ng.: negated, bot.: bottom, incl.: inclusion, asser.: assertion.}  \label{semdlss}
\end{longtable}}


Let $\mathcal{R}$, $\mathcal{T}$, and $\mathcal{A}$  be as above. An interpretation $\I= (\Delta ^ \I, \Delta_{\D}, \cdot ^ \I)$ is a $\D$-model of $\mathcal{R}$ (resp., $\mathcal{T}$), and we write $\I \models_{\D} \mathcal{R}$ (resp., $\I \models_{\D} \mathcal{T}$), if $\I$ satisfies each axiom in $\mathcal{R}$ (resp., $\mathcal{T}$) according to the semantic rules in Table \ref{semdlss}.  Analogously,  $\I= (\Delta^ \I, \Delta_{\D}, \cdot^\I)$ is a $\D$-model of $\mathcal{A}$, and we write $\I \models_{\D} \mathcal{A}$, if $\I$ satisfies each assertion in $\mathcal{A}$, according to the semantic rules in Table \ref{semdlss}. 

A $\shdlssx$-knowledge base $\mathcal{K}=(\mathcal{A}, \mathcal{T}, \mathcal{R})$ is consistent if there is an interpretation $\I= (\Delta^ \I, \Delta_{\D}, \cdot^\I)$ that is a $\D$-model of $\mathcal{A}$,  $\mathcal{T}$, and $\mathcal{R}$.

%
%
Some considerations on the expressive power of $\shdlssx$ are in order. As illustrated in Table \ref{semdlss} existential quantification is admitted only on the left hand side of inclusion axioms. As mentioned in the Introduction, $\shdlssx$ is less powerful than logics such as $\sroiqd\space$ \cite{Horrocks2006}
for what concerns the generation of new individuals. On the other hand, $\shdlssx$ is more liberal than $\sroiqd\space$ in the definition of role inclusion axioms since roles involved are not required to be subject to any ordering relationship, and the notion of simple role is not needed.
For example, the role hierarchy presented in \cite[page~ 2]{Horrocks2006} is not expressible in $\sroiqd\space$ but can be represented in $\shdlssx$. In addition, $\shdlssx$ is a powerful rule language
able to express rules with negated atoms such as
\\  \centerline{$Person(?p) \wedge \neg hasHome(?p, ?h) \implies HomelessPerson(?p)$.}\\ Notice that rules with negated atoms are not supported by the SWRL language.

\section{Overview of the reasoner}
In this section we provide both a general overview and some technical details of the reasoner under implementation. 

\begin{figure}[h] 
	\centering
	\includegraphics[scale=0.47]{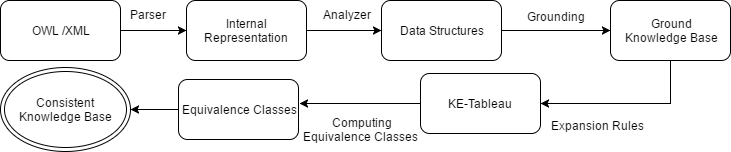} \caption{Execution cycle of the reasoner.} \label{FigOver}
\end{figure}

The input of the reasoner is an OWL ontology serialized in the OWL/XML syntax (see Figure \ref{FigOver}). 

If the ontology meets the $\shdlssx$ requirements, then a parser produces the internal coding of all axioms and assertions of the ontology in set-theoretic terms as a list of strings. Such translation exploits the function $\theta$ used in \cite{RR2017} to map $\shdlssx$-KBs to $\flqsr$-formulae. Each such string represents either a $\flqsr$-quantifier free formula or a $\flqsr$ purely universally quantified formula whose quantifiers have been moved as inward as possible. 
In the subsequent step, the reasoner builds the data-structures required to execute the algorithm, then it constructs the
expansion of each $\flqsr$ purely universally quantified formula according to \cite[page 9]{RR2017} yielding an expanded (ground) $\KB$, $\Phi_\KB$.  Then a \ke\space $\T_\KB$, representing the saturation of $\KB$, is constructed.  

Let $\Phi \defAs \{ C_1,\ldots, C_p\}$ be a collection of disjunctions of $\flqsr$-quantifier free atomic formulae of level $0$ of the types: $x =y$, $x \in X^1$, $\langle x,y\rangle \in X^3$, $\neg (x =y)$, $\neg (x \in X^1)$, $\neg(\langle x,y\rangle \in X^3)$. $\mathcal{T}$ is a \textit{\ke}\space for $\Phi$ if there exists a finite sequence $\mathcal{T}_1, \ldots, \mathcal{T}_t$ of trees such that (i) $\mathcal{T}_1$ is a one-branch tree consisting of the sequence $C_1,\ldots, C_p$, (ii) $\mathcal{T}_t = \mathcal{T}$, and (iii) for each $i<t$, $\mathcal{T}_{i+1}$ is obtained from $\mathcal{T}_i$ either by an application of one of the  rules in Fig. \ref{exprule} or by applying a substitution $\sigma$ to a branch $\vartheta$ of $\mathcal{T}_i$ (in particular, the substitution $\sigma$ is applied to each atomic formula $X$ of $\vartheta$; the resulting branch will be denoted by $\vartheta\sigma$). The set of atomic formulae $\seq \defAs \{ \overline{\beta}_1,\ldots,\overline{\beta}_n\} \setminus \{\overline{\beta}_i\}$ occurring as premise in the E-rule contains the complements of all the components of the formula $\beta$ with the exception of the component $\beta_i$.  

{
	\begin{figure}[h]
		{{
				\begin{center}
					\begin{minipage}[h]{5cm}
						$\infer[~~~\textbf{E-Rule}]
						{\beta_i}{\beta_1 \vee \ldots \vee \beta_n & \quad \seq}$\\[.1cm]
						{ where $\seq \defAs \{ \overline{\beta}_1,...,\overline{\beta}_n\} \setminus \{\overline{\beta}_i\}$,}\\[-.1cm] { for $i=1,...,n$}
					\end{minipage}~~~~~~~~~~~
					\begin{minipage}[h]{2.5cm}
						\vspace{0.1cm}
						$\infer[~~~\textbf{PB-Rule}]
						{A~~|~~\overline{A}}{}$\\[.1cm]
						{ with $A$ an atomic formula}
					\end{minipage}
				\end{center}
				\vspace{-.2cm}
			}
			\caption{\label{exprule}Expansion rules for the \ke.}
		}
	\end{figure}
}

Let $\mathcal{T}$ be a \ke. A branch $\vartheta$ of $\mathcal{T}$  is \textit{closed} if it contains either both $A$ and $\neg A$, for some atomic formula $A$, or an atomic formula of type $\neg(x = x)$. Otherwise, the branch is \textit{open}. A \ke\space is \emph{closed} if all its branches are closed. A formula $\beta_1 \vee \ldots \vee \beta_n$ is \textit{fulfilled} in a branch $\vartheta$, if $\beta_i$ is in $\vartheta$, for some $i=1,\ldots,n$; otherwise it is \emph{unfulfilled}. A branch $\vartheta$ is \textit{fulfilled} if every formula $\beta_1 \vee \ldots \vee \beta_n$ occurring in $\vartheta$ is fulfilled; otherwise it is \emph{unfulfilled}. 
A branch $\vartheta$ is \textit{complete} if either it is closed or it is open, fulfilled, and it does not contain any atomic formula of type $x=y$, with $x$, $y$  distinct variables. A \ke\space is \textit{complete} (resp., \emph{fulfilled}) if all its  branches are complete (resp., fulfilled or closed).  

Procedure \textsf{saturate}-$\shdlssx$-$\KB$ is illustrated in Figure \ref{proc1}.





{
	\begin{figure}
		
	\begin{algorithmic}[1]\Procedure{\textsf{saturate}-$\shdlssx$-$\KB$}{$\phi_\KB$};
		
		
		\State - let $\Phi_\KB$ be the expansion of $\phi_\KB$ 
		;
		
		\State $\T_{\KB}$ := $\Phi_\KB$;
		\While{$\T_{\KB}$ is not fulfilled}
		\parState{- select an unfulfilled open branch $\vartheta$ of $\T_{\KB}$ and an unfulfilled formula\\ \hspace*{.2cm}$\beta_1 \vee \ldots \vee \beta_n$ in $\vartheta$;}
		\If{$\seqnj$ is in $\vartheta$, for some $j \in \{1,\ldots,n\}$}
		\State - apply the E-Rule to $\beta_1 \vee \ldots \vee \beta_n$ and $\seqnj$ on $\vartheta$;
		\Else 
		\parState{- let $B^{\overline{\beta}}$ be the collection of the atomic formulae $\overline{\beta}_1,\ldots,\overline{\beta}_n$ present in $\vartheta$
			and let\\ \hspace*{.15cm} $h$ be the lowest index such that $\overline{\beta}_h \notin B^{\overline{\beta}}$;}
		\parState{- apply the PB-rule to $\overline{\beta}_h$ on $\vartheta$;}
		\EndIf;
		\EndWhile;
		\parWhile{$\T_\KB$ has open branches containing atomic formulae of type $x = y$, with distinct $x$ and $y$}
		\State{- select such an open branch $\vartheta$ of $\T_\KB$;}
		
		\State $\sigma_{\vartheta} := \epsilon$ (where $\epsilon$ is the empty substitution);
		
		\State $\mathsf{Eq}_{\vartheta} := \{ \mbox{atomic formulae of type $x = y$ occurring in $\vartheta$}\}$;
		
		\While{$\mathsf{Eq}_{\vartheta}$ contains $x = y$, with distinct $x$, $y$}
		
		\State - select an atomic formula $x = y$ in $\mathsf{Eq}_{\vartheta}$, with distinct $x$, $y$;
		
		\State $z := \min_{<_{\vartheta}}(x,y)$;
		
		\State $\sigma_{\vartheta} := \sigma_{\vartheta} \cdot \{x/z, y/z\}$;
		
		\State $\mathsf{Eq}_{\vartheta} := \mathsf{Eq}_{\vartheta}\sigma_{\vartheta}$;
		\EndWhile;
		\EndparWhile;
		\State \Return $(\T_\KB)$;
		\EndProcedure;
		
	\end{algorithmic}
     \caption{\ke\space procedure for the saturation of $\shdlssx$ KBs.}
     \label{proc1}
     \end{figure}
}

Initially a one-branch \ke\space $\T_{\KB}$ for $\Phi_\KB$ is constructed. Then, $\T_{\KB}$ is expanded by systematically applying 
the E-Rule (elimination rule) and the PB-Rule (principle of bivalence rule) in Figure \ref{exprule} to formulae of type $\beta_1 \vee \ldots \vee \beta_n$ till saturation, giving priority to the application of the E-Rule. Once such rules are no longer applicable, for each open branch $\vartheta$ of the resulting \ke, atomic formulae of type $x = y$ occurring in $\vartheta$ are treated by storing in $\vartheta$ 
the equivalence class of $x$ and $y$.

\subsection{Some implementation details}

We first show how the internal coding of $\shdlssx$-KBs represented in terms of $\flqsr$ is defined and how data-structures for the representation of formulae, nodes, and KE-tableaux are implemented. Then we describe the most relevant functions that implement the algorithm. 

$\flqsr$ elements are mapped into string as follows. Variables of type $X^i_{\mathit{name}}$ are mapped into strings of the form $\mathit{Vi}\{\mathit{name}\}$.\footnote{For the sake of uniformity, variables of sort 0 are denoted with $X^0, Y^0, \ldots$. We recall that an individual $a$, a concept $C$, and a role $R$ of a $\shdlssx$-KB are  respectively mapped into the variables $X^0_a$, $X^1_C$, and $X^3_R$, by the function $\theta$ described in \cite{RR2017}.} 
The symbols $\forall$, $\wedge$, $\vee$, $\neg \wedge$, $\neg \vee$ are mapped into the strings \texttt{\$FA}, \texttt{\$AD}, \texttt{\$OR}, \texttt{\$DA}, \texttt{\$RO}, respectively. The relators $\in$, $\not \in$, $=$, $\neq$ are mapped into the strings \texttt{\$IN}, \texttt{\$NI}, \texttt{\$EQ}, \texttt{\$QE}, respectively. A pair $\langle X^0_1,X^0_2\rangle$ is mapped in the string \texttt{ \$OA\, V0{1}\, \$CO \,V0{2}\, \$AO}, where \texttt{\$OA} represents the bracket \textquotedblleft $\langle$'', \texttt{\$AO} the bracket \textquotedblleft {$\rangle$}", and \texttt{\$CO} the comma symbol.


$\flqsr$ variables are implemented by means of the class \texttt{Var} that has three fields. The field \texttt{type} of type integer  defines the sort of the $\flqsr$ variables, the field \texttt{name} of type string represents the name of the variable, and  the field \texttt{var} of type integer set to 0 in case of free variables and to 1 in case of purely universally quantified (bound) variables.

Purely universally quantified variables and free variables are collected  in the vectors  \texttt{VQL} and \texttt{VVL} respectively, that provide a subvector for each sort of variable. The access to \texttt{VQL} and \texttt{VVL} is masked by the class \texttt{VariableSet}.

The operators admitted in $\flqsr$ and internally coded as strings are mapped in three vectors that are fields of the class \texttt{Operator}. Specifically, we identify the vector  \texttt{boolOp}  with values \texttt{\$OR}, \texttt{\$AD}, \texttt{\$RO}, \texttt{\$DA},  the vector \texttt{setOp} with values \texttt{\$IN}, \texttt{\$EQ}, \texttt{\$NI}, \texttt{\$QE}, \texttt{\$OA}, \texttt{\$AO}, \texttt{\$CO}, and the vector \texttt{qutOp} with values \texttt{\$FA}. 

%
%

$\flqsr$ atomic formulae are stored using the class \texttt{Atom} that has two fields. The field \texttt{atomOp} of type integer represents the operator of the formula and corresponds to the index of one of the first four elements of the vector \texttt{setOp}. The field \texttt{components} is a vector whose elements point to the variables 
involved in the atomic formula and stored in \texttt{VQL} and \texttt{VVL}. 


$\flqsr$ formulae are represented by the class \texttt{Formula} having a binary tree-shaped structure, whose nodes contain an object of the class \texttt{Atom}.
The left and the right children contain the left subformula and the right subformula, respectively. The class \texttt{Formula} contains the following fields. The field \texttt{atom} of type pointer to \texttt{Atom} represents the atomic formula. The field \texttt{operand} of type integer represents the propositional operator and his value is the index of the corresponding element of the vector \texttt{boolOp}. 
The field \texttt{psubformula} of type pointer to \texttt{Formula} is the pointer to the father node, while the field \texttt{lsubformula}  and the field \texttt{rsubformula}  contain the pointers to the
nodes representing the left and the right component  of the formula, respectively.


The \ke\space decision procedure is based on the data-structure implemented by the class \texttt{Tableau}. This class uses the instances of the class \texttt{Node} that represents the nodes of the \ke. The class \texttt{Node} has a tree-shaped structure and four fields,  the field \texttt{setFormula} of type vector of  \texttt{Formula} that collects the formulae of the current node, and three pointers to instances of the class \texttt{Node}. These fields are called \texttt{leftchild}, \texttt{rightchild} and \texttt{father} and point to the left child node, to the right child node, and to the father node, respectively. For the root node, the field \texttt{father} is set to NULL. 

Concerning the class \texttt{Tableau}, the root node contains the field \texttt{root} of type pointer to \texttt{Node}. The set of open branches is collected in the field  \texttt{openbranches}, while the  set of closed branches is maintained in the vector called \texttt{closedbranches}. In addition, the class \texttt{Tableau} is provided with the field \texttt{EqSet} that is a three-dimensional vector of integers storing the equivalence classes induced by atomic formulae of type $X^0 = Y^0$, for each branch $\theta$ of the tableau and for each variable of $\theta$ occurring in an atomic formula of type $X^0 = Y^0$.
%

As stated above, the first step of the reasoner consists in parsing the ontology from the OWL/XML file. Such a task is performed by the function \texttt{readOWLXML}
that takes in input the string obtained by reading the OWL/XML file and returns a vector of strings representing the internal coding of the KB. The elements of the obtained vector are analysed and parsed by the function \texttt{insertFormulaKB}
that 
returns an object of type \texttt{Formula} representing the input formula. The function \texttt{insertFormulaKB} builds also the vectors \texttt{VVL} and \texttt{VQL}. 

Once all input formulae have been parsed, the reasoner constructs the expansion of the KB by means of the procedure \texttt{expandKB} that yields the vector of the output formulae (\texttt{out}) from the vector of the input formulae (\texttt{inpf}). 
In order to instantiate all the quantified variables, \texttt{expandKB} exploits a stack and the vectors \texttt{VVL} and \texttt{VQL}. After this step, the reasoner checks for atomic clashes in the expanded KB by means of the procedure \texttt{checkNodeClash}.

The construction of the \ke\space is performed by procedure \texttt{expandTableau} that exploits two stacks of type vector of pointers to \texttt{Node}. The first stack, namely \texttt{noncomBranches}, keeps track of the non-complete branches, while the second one, called \texttt{unfulFormula}, keeps track of the unfulfilled disjunctive formulae.  Initially,  \texttt{expandTableau} attempts to empty the stack \texttt{unfulFormula} by selecting iteratively its elements  and applying  either the  procedure \texttt{ERule}
or   the procedure \texttt{PBRule}, respectively implementing 
the E-Rule and the PB-Rule described in Figure \ref{exprule}, according to procedure \textsf{saturate}-$\shdlssx$-$\KB$ in Figure \ref{proc1}.
The disjuncts of the current formula are stored in a temporary vector and selected iteratively. If a disjunct has its negation on the branch, it is removed from the vector. Once all disjuncts of the formula have been selected, if there is only  an element in the stack, then the procedure \texttt{ERule} is applied to the disjunctive formula. If there is more than one element in the vector, then the procedure \texttt{PBRule} is applied. In case the stack is empty, a contradiction is found and the branch is closed. Clash checks are performed at each insertion of formula, and if a branch gets closed, it is added to 
\texttt{closedbranches}. 

The procedure \texttt{expandTableau}
terminates when either  \texttt{noncomBranches} or \texttt{unfulFormula} are empty. When the procedure terminates with some element in \texttt{noncomBranches}, such branches are added to the vector  \texttt{openbranches}.

The subsequent phase consists in constructing the set of equivalence classes \texttt{EqSet} for each open branch computed by \texttt{expandTableau}.

\texttt{EqSet} is computed by the procedure \texttt{computeEqT}.
For each open branch in \texttt{openbranches}, the procedure searches for formulae of type $X^0 = Y^0$, where $X^0$ and $Y^0$ are selected with respect to the ordering provided by the vector \texttt{VVL}, and stores in \texttt{EqSet} the equivalence class for each variable.  

The procedure terminates when all open branches of the vector \texttt{openbranches} have been analysed. Then, \texttt{EqSet} is used to check for clashes. Finally, if  the vector \texttt{openbranches} is not empty the KB is returned as consistent. 



\subsection{Example of reasoning in $\shdlssx$ }

\begin{figure}[H] 
\centering\includegraphics[scale=0.8]{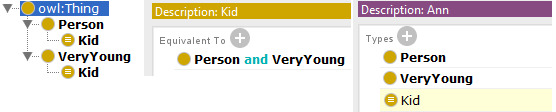}
\caption{A simple OWL ontology.}
\label{imgOWL}
\end{figure}

In this section we show an example of reasoning in $\shdlssx$ and the results provided by the reasoner. For space limitations, we consider the simple OWL ontology illustrated in Figure \ref{imgOWL}.

Then its representation in terms of the description logic $\shdlssx$ is: 
\begin{align*} 
\KB  = (&\{\  \mathsf{Kid} \equiv \mathsf{Person} \sqcap \mathsf{VeryYoung} \},\\
&\{ \mathsf{Person}(Ann)\}
\end{align*} %


As mentioned above, the  mapping function $\theta$ (cfr.  \cite{RR2017}) is applied to $\KB$ thus yielding the following $\flqsr$ representation $\phi_\KB$ of $\KB$. 

\begin{align*} 
\phi_\KB &= (\forall x)((\neg(x \in X_{\mathsf{Kid}}^1) \vee x \in X_{\mathsf{Person}}^1)\\ &\qquad \quad \wedge(\neg (x \in X_{\mathsf{Kid}}^1) \vee x \in X_{\mathsf{VeryYoung}}^1)\\ &\qquad \quad \wedge 
  (\neg (x \in X_{\mathsf{Person}}^1) \vee \neg(x \in  
 X_{\mathsf{VeryYoung}}^1) \vee x \in X_{\mathsf{Kid}}^1))\\
 &\qquad \quad \wedge x_{Ann} \in X_{\mathsf{Person}}^1\,. 
\end{align*} %

Then $\phi_\KB$ is converted in conjunctive normal form, universal quantifiers are moved as inward as possible, and universally quantified variables are renamed so as to make them pairwise distinct. The resulting $\flqsr$-formula $\bar{\phi}_\KB$ is shown in what follows. 

\begin{align*} 
\bar{\phi}_\KB = &(\forall x)((\neg(x \in X_{\mathsf{Kid}}^1) \vee x \in X_{\mathsf{Person}}^1) \wedge\\
 &(\forall y) (\neg (y \in X_{\mathsf{Kid}}^1) \vee y \in X_{\mathsf{VeryYoung}}^1) \wedge \\ 
 &(\forall z) (\neg (z \in X_{\mathsf{Person}}^1) \vee \neg(z \in  
 X_{\mathsf{VeryYoung}}^1) \vee z \in X_{\mathsf{Kid}}^1))\wedge\\
 & x_{Ann} \in X_{\mathsf{Person}}^1 \, .
\end{align*} 

\begin{figure}[H]
\centering
\includegraphics[scale=0.7]{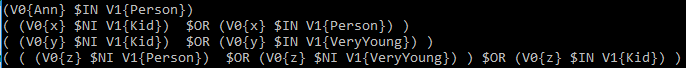}
\caption{Internal representation of the $\KB$.}
\label{imgIntRep}
\end{figure}

\begin{figure}[H]
\centering
\includegraphics[]{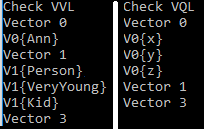}
\caption{The vectors \texttt{VVL} and \texttt{VQL} of the $\KB$.}
\label{imgvvl}
\end{figure}

The internal representation of $\bar{\phi}_\KB$ computed by the reasoner is illustrated in Figure \ref{imgIntRep}, while the vectors \texttt{VVL} and \texttt{VQL} in Figure \ref{imgvvl}.

Then the expansion $\Phi_{\KB}$ of $\bar{\phi}_\KB$ is computed, consisting in the collection of disjunctions of $\flqsr$-quantifier-free atomic formulae of level 0 illustrated in the following. 

\begin{align*} 
\Phi_{\KB} = &\{\neg(x_{A} \in X_{\mathsf{Kid}}^1) \vee x_{Ann} \in X_{\mathsf{Person}}^1,\\
&\ \neg (x_{A} \in X_{\mathsf{Kid}}^1) \vee x_{A} \in X_{\mathsf{VeryYoung}}^1,\\
&\ \neg (x_{A} \in X_{\mathsf{Person}}^1) \vee \neg(x_{A} \in  
 X_{\mathsf{VeryYoung}}^1) \vee x_{Ann} \in X_{\mathsf{Kid}}^1,\\ 
 &\ x_{Ann} \in X_{\mathsf{Person}}^1 \,.
\end{align*}

\begin{figure}[H]
\centering
\includegraphics[scale=0.6]{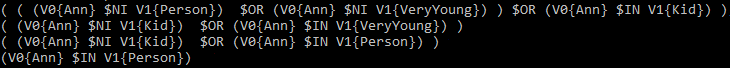}
\caption{Expansion of $\KB$ as computed by the reasoner.}
\label{imgExp}
\end{figure}

The reasoner computes $\Phi_{\KB}$ by means of the function \texttt{expandKB} yielding the result shown in Figure \ref{imgExp}, where each line of the console output is the internal representation of an object of type \texttt{Formula}. According to the procedure of Figure \ref{proc1}, the initial \ke\space $\mathcal{T}_\KB$ computed  by the expansion function \texttt{expandKB} is constituted by the set of formulae $\Phi_{\KB}$. Specifically,  $\Phi_{\KB}$ is stored in the field \texttt{setFormula} of a object \texttt{Node}, that is the initial node of the class \texttt{Tableau}.

\begin{figure}[H]
\centering
\includegraphics[scale=0.6]{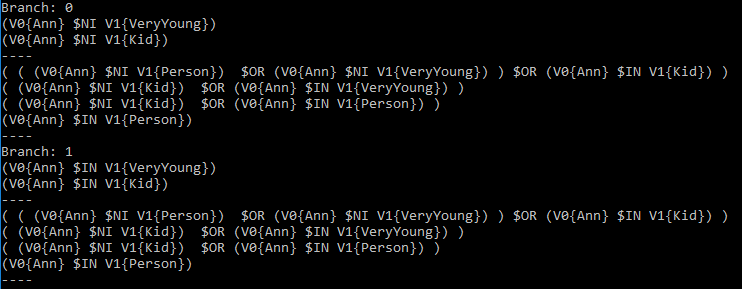}
\caption{The open branch of the \ke\space computed by the reasoner.}
\label{imgOpenBranch}
\end{figure}

Then $\T_{\KB}$ is expanded by systematically applying the E-Rule and the PB-Rule in Figure \ref{exprule} to formulae of type $\beta_1 \vee \ldots \vee \beta_n$ till all $\beta$-formulae have been analysed. The final \ke\space that consists of two complete open branches is illustrated in Figure \ref{imgTableau}.  The complete open branches computed by the reasoner are shown  in Figure \ref{imgOpenBranch}. 

\begin{figure}[H]
\centering

\begin{tikzpicture}[auto, node distance=0.9cm]

\node (A) { $\neg(x_{Ann} \in X_{\mathsf{Kid}}^1) \vee x_{Ann} \in X_{\mathsf{Person}}^1$};
\node (B) [below of=A] {$\neg (x_{Ann} \in X_{\mathsf{Kid}}^1) \vee x_{Ann} \in X_{\mathsf{VeryYoung}}^1$};
\node (C) [below of=B] {$\neg (x_{Ann} \in X_{\mathsf{Person}}^1) \vee \neg(x_{Ann} \in  
	X_{\mathsf{VeryYoung}}^1) \vee x_{Ann} \in X_{\mathsf{Kid}}^1$};
\node (D) [below of=C] {$ x_{Ann} \in X_{\mathsf{Person}}^1$}; 
\node (F) [label={[label distance=-1.3cm]}, below left of=D,left=1.5cm, below=0.6cm] {$x_{Ann} \in X_{\mathsf{VeryYoung}}^1$};
\node (G) [below right of=D, right=1.5cm,below=0.6cm] {$\neg(x_{Ann} \in X_{\mathsf{VeryYoung}}^1)$};
\node (H) [below of=F, below=0.3cm, label={[label distance=-1.3cm] \color{green}{Complete}} ]{$x_{Ann} \in X_{\mathsf{Kid}}^1$};
\node (I) [below of=G, below=0.3cm, label={[label distance=-1.3cm] \color{green}{Complete}} ]{$\neg(x_{Ann} \in X_{\mathsf{Kid}}^1)$};

\path (A) edge  (B);
\path (B) edge (C);
\path (C) edge (D);

\path (D) edge  node [below=0.1cm, right=0.4cm]{\texttt{PB-Rule}}  (F)
edge (G);
\path (F) edge node [below=0.1cm, right=0.4cm]{\texttt{E-Rule}} (H);          
\path (G) edge node [below=0.1cm, left=0.4cm]{\texttt{E-Rule}}(I);

\end{tikzpicture}
\caption{\ke\space for $\Phi_{\KB}$.}
\label{imgTableau}

\end{figure}

\begin{figure}[H] \centering
\begin{minipage}[b]{.4\textwidth}
\centering
\includegraphics[]{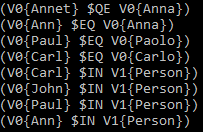}
\caption{The one-branch \ke\space $\T_{\KB_2}$.}
\label{EQ1}
\end{minipage}%
\hfill
\begin{minipage}[b]{.4\textwidth}
\centering
\includegraphics[]{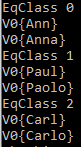}
\caption{Set of equivalence classes for $\T_{\KB_2}$.}
\label{EQ2}
\end{minipage}

\end{figure}

In the last step, the reasoner computes for each open complete branch the equivalent classes for the individuals involved in formulae of type $x=y$ and checks for inconsistency. Consider the following knowledge base $\KB_2$. $\T_{\KB_2}$ is the consistent one-branch \ke\space shown in Figure \ref{EQ1}.

\begin{align*}
\KB_2 = (\{ &Person(Ann), \, Person(Paul),\, Person(John),\, Person(Carl),\,\\
           &Annet \neq Ann,\, Ann=Anna,\, Paul= Paolo,\, Carl=Carlo\})
\end{align*}

For the single branch of $\T_{\KB_2}$, the three equivalence classes computed by the reasoner are shown in Figure \ref{EQ2}.

\section{Conclusions}
We have presented an ongoing implementation of a \ke\space based decision procedure for the consistency problem of $\shdlssx$-KBs in terms of set-theoretical
$\flqsr$-formulae.  
The reasoner, developed in \textsf{C++}, takes as input
OWL ontologies serialized in the OWL/XML format.

Currently, the tool is in its beta-testing phase. We plan to compare it with existing reasoners such as Hermit \cite{ghmsw14HermiT} and Pellet \cite{PelletSirinPGKK07}, and to provide some benchmarking. Then, we intend to extend the reasoner with the HOCQA procedure \cite{RR2017}, thus providing ABox reasoning services. We also plan to allow data type reasoning by integrating  
Satisfiability Modulo Theories solvers. 
Moreover, techniques developed in \cite{CanNicOrl10,CanNicOrl11} will be used to include reasoning for description logics admitting full existential and universal restrictions.  Finally, we intend to implement a parallel version of the software by exploiting Message Passing Interface, since each branch of the \ke\space can be computed by a single processing unit.

\bibliographystyle{plain}
\bibliography{biblio}

\end{document}